\documentclass[twocolumn,prx,final,showpacs,amsmath,amssymb,superscriptaddress,longbibliography]{revtex4-1}
\usepackage[toc,page]{appendix}
\usepackage{multirow}
\usepackage{amsmath,amssymb}
\usepackage{graphicx,float}
\usepackage{times,txfonts}
\usepackage{color}
\usepackage{multirow}
\usepackage{bbold}
\usepackage{color}
\usepackage{soul}
\usepackage{hyperref}
\usepackage{times,txfonts}
\usepackage{natbib}
\usepackage{nicefrac}
\usepackage{blindtext}
\usepackage[export]{adjustbox}

\newcommand{\qo}[1]{``#1''}

\newcommand{\etal}{\textit{et~al.}}
\renewcommand{\epsilon}{\varepsilon}
\renewcommand{\phi}{\varphi}

\usepackage{sistyle}
\usepackage{soul}
\definecolor{lightblue}{RGB}{185,210,248}
\definecolor{YellowOrange}{RGB}{226,154,2}
\sethlcolor{YellowOrange}
\sethlcolor{lightblue}

\begin{document}
\title{Experimental investigation of Popper's proposed ghost-diffraction experiment}
\author{Eliot Bolduc}
\affiliation{Department of Physics, University of Ottawa, 25 Templeton St., Ottawa, Ontario, K1N 6N5 Canada}
\affiliation{SUPA, School of Engineering and Physical Sciences, Heriot-Watt University, Edinburgh EH14 4AS, UK}
\author{Ebrahim Karimi}
\affiliation{Department of Physics, University of Ottawa, 25 Templeton St., Ottawa, Ontario, K1N 6N5 Canada}
\affiliation{Department of Physics, Institute for Advanced Studies in Basic Sciences, 45137-66731 Zanjan, Iran}
\author{Kevin Pich\'e}
\affiliation{Department of Physics, University of Ottawa, 25 Templeton St., Ottawa, Ontario, K1N 6N5 Canada}
\author{Jonathan Leach}
\affiliation{SUPA, School of Engineering and Physical Sciences, Heriot-Watt University, Edinburgh EH14 4AS, UK}
\author{Robert W. Boyd}
\affiliation{Department of Physics, University of Ottawa, 25 Templeton St., Ottawa, Ontario, K1N 6N5 Canada}
\affiliation{Institute of Optics, University of Rochester, Rochester, New York, 14627, USA}
\begin{abstract}
In an effort to challenge the Copenhagen interpretation of quantum mechanics, Karl Popper proposed an experiment involving spatially separated entangled particles. In this experiment, one of the particles passes through a very narrow slit, and thereby its position becomes well-defined. This particle therefore diffracts into a large divergence angle; this effect can be understood as a consequence of the Heisenberg uncertainty principle. Popper further argued that its entangled partner would become comparably localized in position, and that, according to his understanding of the Copenhagen interpretation of quantum mechanics, the \qo{mere knowledge} of the position of this particle would cause it also to diffract into a large divergence angle. Popper recognized that such behaviour could violate the principle of causality in that the slit could be removed and the partner particle would be expected to respond instantaneously. Popper thus concluded that it was most likely the case that in an actual experiment the partner photon would not undergo increased diffractive spreading and thus that the Copenhagen interpretation is incorrect. Here, we report and analyze the results of an implementation of Popper's proposal. We find that the partner beam does not undergo increased diffractive spreading. Our work resolves many of the open questions involving Popper's proposal, and it provides further insight into the nature of entanglement and its relation to the uncertainty principle of correlated particles.
\end{abstract}
\maketitle

\section{Introduction} 
Among different theories of physics, quantum mechanics plays an essential role in describing atomic and subatomic phenomena. Despite the fact that its formalism passes every rigorous experimental test, quantum mechanics has lead to many controversies since its inception. In a seminal paper, Einstein, Podolsky and Rosen (EPR) conceived a \emph{gedankenexperiment} using entangled particles to argue against the completeness of quantum mechanics~\cite{epr:35}. They made use of a pair of particles that are assumed perfectly correlated in both position and momentum - the EPR state.  Their concern was that their thought experiment appeared to allow for the simultaneous reality of conjugate quantities, in apparent conflict with the Heisenberg uncertainty principle (HUP). The modern answer to the EPR argument is that quantum mechanics is a \emph{truly} nonlocal theory~\cite{bell:64}, although it is nonetheless consistent with causality in that information cannot be transmitted faster than the speed of light, i.e.~it obeys a no-signaling principle~\cite{ghirardi:88}. 

\begin{figure}[h]
	\centering
	\includegraphics[width = 0.47 \textwidth]{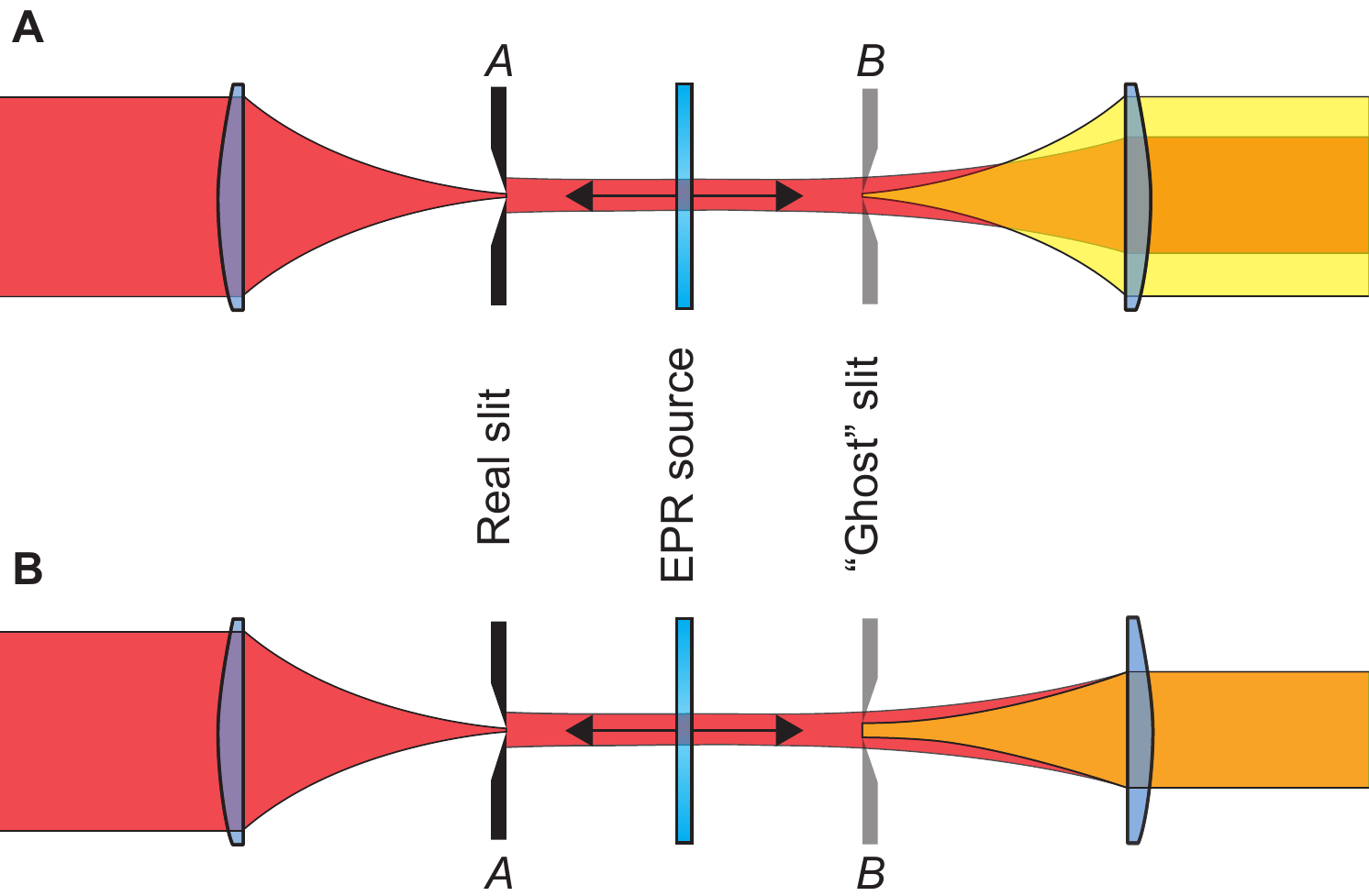} 
	\caption{\label{fig:schematic} {\bf Different outcomes of Popper's experiment.} Popper assumes a source (labeled EPR) of two particles that are entangled in both position and transverse momentum.    If the particle on the left passes through a narrow slit at position {\emph A} (as could be determined by placing a large area or \qo{bucket} detector after the slit), the position of the particle (in the vertical dimension)  becomes known to high accuracy, and, by the HUP, its momentum becomes highly uncertain.  The particle thus diffracts into a large divergence angle.  The question asked by Popper was whether this particle's entangled partner moving to the right would also undergo increased diffractive spreading, because its position is also very well defined at the position of the \qo{ghost} slit. Popper perceives two possible outcomes, as shown in parts {\bf A} and {\bf B}.  
In each part of the figure we use the following colour coding: red indicates that a detector will measure a signal from the source (this is the full two-photon field); orange indicates the overlap of the red region and the region where one would obtain a coincidence count conditioned on the detector located behind the slit on left-hand side firing; and yellow indicates the the region where one would obtain a conditional coincidence count but outside of the support of the red region.
In the first scenario ({\bf A}), the width of the diffraction arising from the  ghost slit, as measured  in coincidence, is comparable to the width of the diffraction that would arise from a real slit, indicated by the wide red shading on side \emph{A}. This result is incompatible with causality in that placing a real slit on side \emph{A} would instantaneously influence the diffraction spread on side \emph{B}. In the second scenario ({\bf B}), the beam on the right as measured in coincidence becomes localized in the plane of the ghost slit (as this is a normal feature of entanglement) but does not undergo  increased spreading in the far-field of the slit.  According to Popper's understanding, the Copenhagen interpretation of quantum mechanics predicts the first scenario, and therefore this theory is incompatible with relativistic causality. Our experimental results confirm the second of Popper's scenarios.}
\end{figure}

In addition to the EPR work, Karl Popper outlined an experiment that challenged the predictions of quantum mechanics based on its Copenhagen interpretation~\cite{popper:34, popper:82}. He designed an experiment that he believed had two possible outcomes, as shown in Fig.~\ref{fig:schematic}. The first outcome would conflict with relativistic causality and the concept that information cannot travel faster than the speed of light, that is, with the no-signaling principle. The second outcome would be in conflict with Popper's understanding of the predictions of the Copenhagen interpretation of quantum mechanics, and specifically with the concept that not all information about a single quantum system can be acquired simultaneously, that is, with the  Heisenberg uncertainty relation between conjugate quantities. Popper favoured causality over the Copenhagen interpretation and tended to favour the second outcome. 

In this article, we present experimental evidence that the second outcome is in fact what is observed, thus confirming that there is no violation of causality. Nonetheless, our observation of the second scenario does not challenge the standard interpretation of quantum mechanics. We present theoretical arguments that show that this result is in fact consistent both with relativistic causality and with the formalism of quantum mechanics.

Since the time of Popper's proposal, many scientists have looked into the questions he raised~\cite{qureshi:05,gerjuoy:06,richardson:12,collett:87,kim:99,short:01,peres:02, ghirardi:07}. Notably, in 1987, Collett and Loudon proposed a more realistic model of Popper's experiment where the source, instead of being a point source, is extended in the transverse direction of space~\cite{collett:87}. They concluded that the transverse uncertainty of the source causes the diffraction spread on side \emph{B} to decrease with the size of the slit on side \emph{A}, hence reconciling quantum mechanics with causality. Ghirardi~\etal~have a radically different view of Popper's experiment~\cite{ghirardi:07}. They claim that the local actions performed at the plane of the slit on side \emph{A} are completely uncorrelated with the diffraction spread on side \emph{B}. In 1999, Kim and Shih~\cite{kim:99} reported  experimental results showing that the \qo{ghost} slit does not induce increased diffractive spreading. However, suspicion with regards to the explanation of their results was raised by Short~\cite{short:01}, who formulated the results of the Popper experiment in terms of conditional uncertainties. Further theoretical work has questioned whether Popper's thought experiment truly addresses either the Copenhagen interpretation or the HUP~\cite{peres:02, ghirardi:07}.
%

\section{Theory}
The thought experiments of EPR and of Popper are closely related to the HUP, one of the fundamental concepts of quantum mechanics. The HUP states that there is a fundamental limit to the accuracy with which one can gain simultaneous knowledge of conjugate quantities, in our case the position and momentum of a single particle as described by $\Delta x\, \Delta p \ge \hbar/2$.  Here $\Delta x$ and $\Delta p$ stand for uncertainty in position and momentum, respectively, and $\hbar$ is the reduced Planck constant.

According to the arguments of EPR, either quantum mechanics is an incomplete theory or else it allows for simultaneous reality of the conjugate quantities position $x$ and momentum $p$ of a quantum system. However, Popper considered a somewhat different situation: he considered the implications of the HUP on the position and momentum of the second particle when in both cases the measurement on the first particle is made on the position degree of freedom. In this situation, one must consider both  position-position and position-momentum correlations. Popper thought that this type of inferred measurement could lead to a violation of the uncertainty principle. Indeed, in the same manner in which the standard uncertainty principle has a lower bound, the uncertainty principle relevant to Popper's thought experiment is~\cite{short:01} 
\begin{align}\label{eq:hup} 
	\Delta (x_B|x_A)\,\Delta (p_B|x_A) \geq \hbar/2.
\end{align}
This equation should be interpreted as the uncertainty product for particle \emph{B} given a measurement of the position of particle \emph{A}.  Below, we show that this inequality can be saturated for the case of a pure two-photon entangled state that exhibits strong EPR correlations.
 
In our implementation of Popper's experiment, the entangled particles are generated through spontaneous parametric downconversion (SPDC),  the theory  of which is well-known~\cite{chan:07,pires:09,walborn:10}.  Here, we consider only the horizontal distribution along $x$, which is perpendicular to the real slit, and sum our two-dimensional data over the $y$ degree of freedom.

At the output of the crystal, the two-photon mode function takes the form~\cite{chan:07,pires:09}
\begin{align}\label{eq:NF} 
	\Psi({x_B},{x_A})=N\,\exp{\left(-\frac{(x_A+x_B)^2}{4 \sigma_p^2}\right)}\,\exp{\left(-\frac{(x_A-x_B)^2}{4\sigma_q^2}\right)},
\end{align}
where $N=(\pi\,\sigma_p \sigma_q)^{-1}$ is a normalization constant, $\sigma_p$ is the $1/e^2$ width of the intensity of the collimated Gaussian-distributed pump, and $\sigma_q$ is the width of the position correlations, which is a function of the length of the crystal (see Materials and Methods for more details). Upon detection of a photon in arm \emph{A} at position $x_A=0$, the conditional state of photon \emph{B} takes the form of a Gaussian distribution: 
\begin{align}\label{eq:NFoversimplified} 
	\left| \Psi({x_B}|x_A=0) \right|^2 \approx N'\, \exp{\left(-{x}_B^2/\left(2\sigma_q^2\right)\right)},
\end{align}
%
where $N'$ is a normalization constant. The standard deviation (SD) of the associated probability distribution is equal to $\Delta(x_B|x_A=0)= \sigma_q$. We used the fact that the pump width $\sigma_p$ ($\approx 450$~$\mu$m) is much greater than that of the mode width of the SPDC position correlations $\sigma_q$ ($\approx 10$~$\mu$m), i.e.~$\sigma_p\gg\sigma_q$. The conditioned mode function of photon \emph{B} can be treated as a pure state (see Eq.~(\ref{eq:NFoversimplified})) and undergoes diffraction like a pure state. Using the Fourier transform of the position wave function of photon \emph{B}, we find the conditional state of photon \emph{B} in momentum space as
\begin{align}\label{eq:FF} 
	\left|\Psi({p_B}|x_A=0)\right|^2\approx N''\, \exp{\left(-2 {p}_B^2\sigma_q^2/\hbar^2 \right)},
\end{align}
where $p_B$ is the momentum of photon \emph{B}, and $N''$ is a normalization constant, respectively. The SD of the probability distribution $|\Psi({p_B}|x_A=0)|^2$ is equal to $\Delta(p_B|x_A=0)=\hbar/(2\sigma_q)$. This distribution is identical to that of the unconditioned Gaussian distribution, and thus, for the case of the entangled state (\ref{eq:NF}), the quantum formalism predicts that scenario {\bf B} from Fig.~\ref{fig:schematic} will occur. 
Finally, taking the product of the conditioned standard deviations associated with Eqs.~(\ref{eq:NFoversimplified}) and (\ref{eq:FF}), we obtain 
\begin{align}\label{eq:HEINSENBERG} 
	\Delta(x_B|x_A=0)\,\Delta(p_B|x_A=0)=\hbar/2,
\end{align}
a result that is consistent with the uncertainty principle in the form of Eq.~(\ref{eq:hup}). This result is to be expected since Eq.~(\ref{eq:NFoversimplified}) and Eq.~(\ref{eq:FF}) are conjugate quantities related by Fourier transforms. It follows that for the initial state~(\ref{eq:NF}), the presence of a slit in \emph{A} does not lead to increased diffractive spreading of photon \emph{B} as the slit width is decreased. Therefore, quantum mechanics leads to the prediction of the second scenario ({\bf B}) shown in Fig.~\ref{fig:schematic}.   

%
\begin{figure}[!htbp]
	\centering
	\includegraphics[width = 0.45 \textwidth]{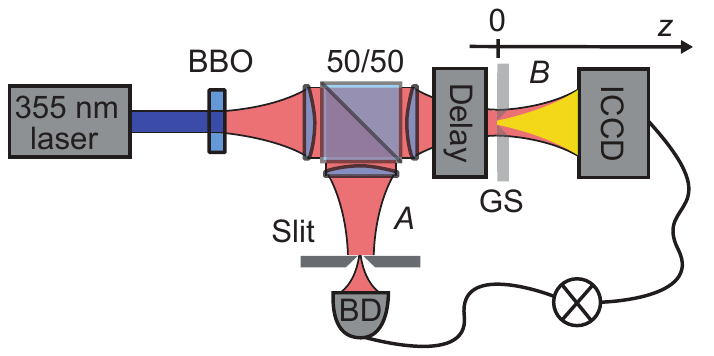} 
	\caption{\label{fig:quantum} \textbf{Schematic of our implementations of Popper's experiment.} Entangled photon pairs are generated through spontaneous parametric downconversion (SPDC) in a $\beta$-barium borate (BBO) crystal using type-I phase matching. The real slit on side \emph{A} is placed in an image plane of the crystal, and a large-area (bucket) detector (BD) is placed after the slit; this detector registers those photons that pass through the slit and thereby determines their position with high accuracy. Photon \emph{B} goes through an image-preserving delay line, and then is captured by an intensified CCD (ICCD) camera that is triggered by the photons that pass through the real slit on side \emph{A} (see Materials and Methods for more details).}
\end{figure}
%
\begin{figure*}
	\centering
	\includegraphics[width=\textwidth]{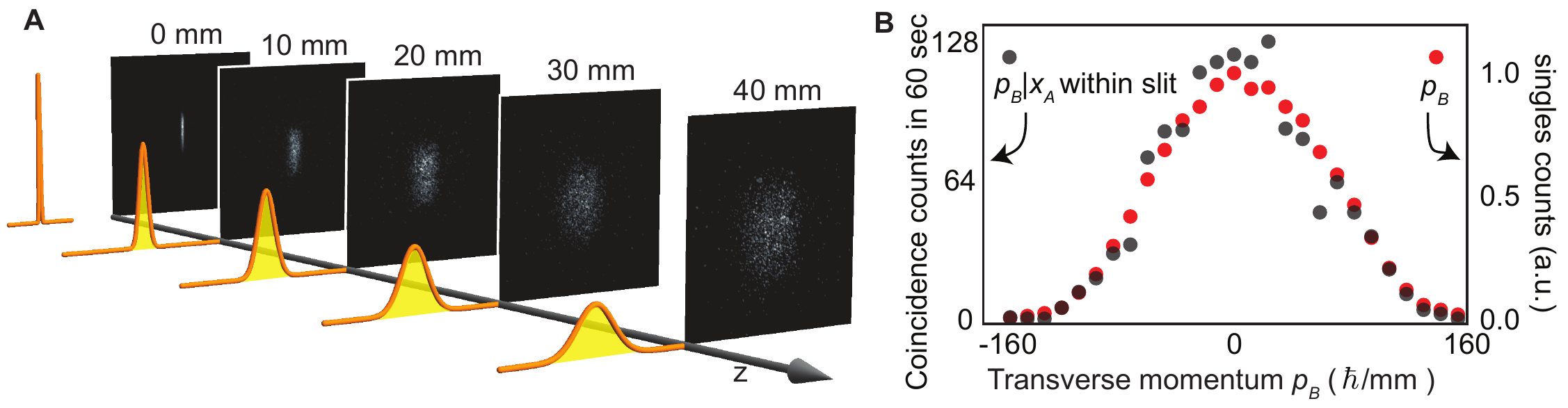} 
\caption{\label{fig:quantumdata} {\bf Experimental observation of  ghost diffraction from a slit.} ({\bf A}) Coincidence images from the ICCD showing  ghost diffraction from a slit upon propagation. Images were recorded between the near-field of the BBO crystal $z=0$~mm and the far-field of the crystal $z\rightarrow\infty$ (not shown). Here we show five images at $10$~mm increments from the  ghost slit. ({\bf B}) Comparison between the  ghost diffraction (conditioned $p_B |x_A$ \text{within slit}) and singles (unconditioned $p_B$) in the far-field ($z\rightarrow\infty$). The black points show data for the conditional case, measured in coincidence; the red points show data for the unconditional case. Note that the conditional and unconditional distribution of photon \emph{B} are essentially identical showing that placing a slit in arm \emph{A} does not influence the momentum distribution of photon \emph{B}.}
\end{figure*}

\section{Experimental Results}
Figure~\ref{fig:quantum} shows the schematic representation of our implemented Popper's thought experiment. Generated photon pairs are split out by means of a $50/50$ non-polarizing beam splitter. They are sent into a 10-$\mu$m-wide slit that located at the image plane of the downconversion crystal in arm \emph{A} and an image-preserving delay line in arm \emph{B}, respectively. Photons that pass through the slit in arm \emph{A} are registered by a bucket detector. This detector registers those photons that pass through the slit and thereby determines their position with high accuracy. We make measurements on a photon in arm \emph{B} that are conditioned on the detection of a photon in arm \emph{A}. 

A triggered intensified CCD camera placed in arm \emph{B} in the image plane of the real slit and thus also in the image plane of the crystal is used to measure in coincidence the width of the ghost slit (see Materials and Methods for more details). In Fig.~\ref{fig:quantumdata}, the frame at $z=0$ mm shows an image of the ghost slit.  The width of this image, expressed in terms of its SD, is $\Delta(x_B|x_A\text{within slit})=(19\pm1)$ $\mu$m. The predicted width of the image is given by the convolution of the transmission function of the real slit  ($10$~$\mu$m wide) with the point-spread function of the imaging system, which is given by the  mode function $|\Psi(x_B,x_A)|^2$ taken to be a function of $x_B$. This calculated width is  about $12$~$\mu$m.  The measured $19$~$\mu$m width of the ghost slit is greater than the calculated width of $12$~$\mu$m because of non-ideal position correlations of the two photons, which we believe arise from misalignment or aberrations in our imaging system. Nonetheless, to within a factor of two, we confirm Popper's expectation that the position of photon \emph{B} is measured with \qo{approximately the same precision} as that of photon \emph{A}. Of course, a complication raises from the fact that width of an image is often specified in terms of its SD. For a uniform transmission distribution, e.g. for a slit of width $d$, this SD is equal $d/\sqrt{12}$. However, in our analysis we take the physical width of the slit to be the measure of the position of photon \emph{A}.

Using a lens with a focal length of 75 mm, we next place the ICCD camera in the far-field (Fourier transform plane) of the ghost slit and record the distributions of the singles and coincidence counts, as shown in Fig.~\ref{fig:quantumdata}-({\bf B}). The standard deviations of these distributions are respectively $\Delta(p_B)/\hbar=(0.048\pm0.001)$ $\mu$m$^{-1}$ and $\Delta(p_B|x_A\text{within slit})/\hbar=(0.046\pm0.006)$ $\mu$m$^{-1}$. The widths of the two distributions are very similar. We conclude from these results that the detection of photon \emph{A} after the real slit does not affect the far-field distribution of photon \emph{B}. To express this thought more quantitatively, we note that for these data we find an uncertainty product of 
 \begin{align}
 \Delta(p_B|x_A \text{within slit})\,\Delta(x_B|x_A \text{within slit})=(0.87\pm0.16)\,\hbar. 
  \end{align} 
This product is greater than $\hbar/2$ and is thus consistent with the uncertainty principle. The measured uncertainty product does not saturate the uncertainty principle of Eq.~(\ref{eq:HEINSENBERG}). As mentioned before, we attribute this increased uncertainty to aberrations and misalignment along the optical axis. 

\begin{figure}
	\centering
	\includegraphics[width=.45\textwidth]{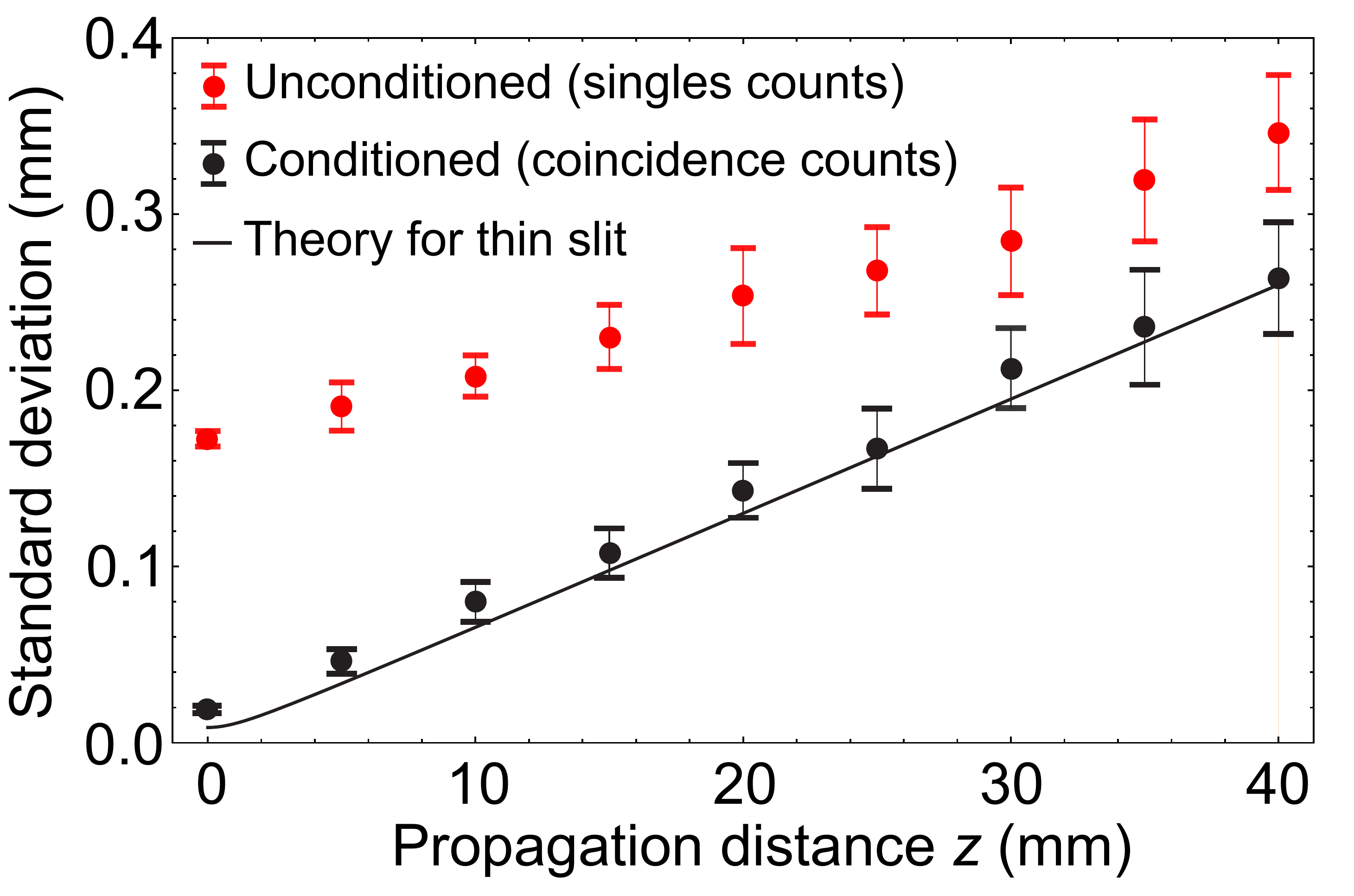} 
\caption{\label{fig:pro} {\bf Measured transverse width of particle B's field as a function of distance from a ghost slit.} The black points show data for the conditional widths; the red points show data for the unconditional widths.  The black line is a fit to the conditioned data using Gaussian beam propagation. The error bars correspond to a 95\% confidence region of the width of a Gaussian distribution fitted to the recorded transverse distributions.}
\end{figure}

In Fig.~\ref{fig:pro}-({\bf A}), we compare the measured width of the conditional diffraction from the ghost slit (black points) to the theoretical width of a propagating Gaussian beam (black curve): $a(z)=a(0)\sqrt{1+(z/z_R)^2}$, where $z_R$ is the Rayleigh range associated with the conditioned mode-function of photon \emph{B}. The theory fits very well for all points other than the point at $z=0$, and we attribute this difference also to a slight misalignment along the optical axis. We also show the width of the singles counts on side \emph{B} (red points) as a function of propagation distance. The width of the singles and the coincidence counts asymptotically approach one another  upon propagation, and as confirmed by Fig.~\ref{fig:quantumdata}-({\bf B}) match perfectly at the far-field of the ghost slit.

%

\section{Discussion}
The results from our experiment confirm that the second scenario outlined in Fig.~\ref{fig:schematic}-({\bf B}) prevails. This outcome is precisely what Popper expected on the basis of the argument that it conforms to the principle of causality. Popper was correct that the events in the near-field of the real slit on side \emph{A} cannot influence the outcome of events in the far field of side \emph{B}~\cite{combourieu:92}. We find that the presence or absence of the real slit in arm \emph{A} leads to no difference between the conditional and unconditional distributions on side \emph{B} in the far field. We can express this thought somewhat differently as follows. The rate of singles counts measured at any position on side \emph{B} is the same whether or not a position measurement is made on side \emph{A}. In contrast, for coincidence measurements the distribution on side \emph{B} is narrower than the singles distribution in the near-field of the ghost slit but has the same width in the far-field. Thus, the presence or absence of the slit on side \emph{A} does not affect the photon distribution in singles or in coincidence in the far field for side \emph{B}. We have also shown theoretically that the quantum formalism is consistent with the second scenario, which is causal. This is in contrast to what Popper thought based on his understanding of the Copenhagen interpretation of quantum mechanics, which would preclude to scenario 2. One possible flaw in Popper's reasoning of the physical situation is that he assumed that the source of particle pairs was both perfectly correlated in transverse momentum and had a very small transverse extent~\cite{collett:87}. Ghirardi~\etal~argue that the laws of quantum mechanics prohibit the simultaneity of these two properties~\cite{ghirardi:07}. Thus, if the position of the source is known very well, then according to \cite{ghirardi:07} no correlations in momentum are possible. 

In the context of Popper's experiment, the relevant uncertainty relation involves uncertainties in the position and the momentum of photon \emph{B} conditioned on the position of photon of \emph{A}~\cite{short:01}. Quantum theory predicts that this product is equal to $\hbar/2$ and saturates the HUP. In our experiment, we obtained a value of $0.87\hbar$ for this product because of imperfections in our laboratory setup. 

\section{Conclusions}
In conclusion, we have implemented Popper's proposed experiment with entangled photon pairs generated through SPDC. Our measurements of the diffraction from the ghost image of a narrow slit closely matches the predictions of the corresponding model, derived from the theory of SPDC. Our results are consistent both with the quantum formalism, i.e. the HUP and with causality (the no-signaling principle) and confirm that there is no spread in the momentum distribution of one entangled photon due to presence of the slit for the other photon. We find that the conditional HUP, $\Delta(x_B|x_A=0)\Delta(p_B|x_A=0)\geq \hbar/2$, is validated theoretically and experimentally.

\section{Acknowledgments:} 
R.W.B. acknowledges the previous discussions with John Howell and Rayan Bennink. This work was supported by the Canada Excellence Research Chairs (CERC) Program. E.B. acknowledges the financial support of the FQRNT, grant number 149713.

\setcounter{figure}{0} \renewcommand{\thefigure}{A\arabic{figure}}
\appendix\footnotesize{
\section{Theory}
In our implementation of Popper's experiment with entangled photons, we used spontaneous parametric downconversion (SPDC) to make a  ghost image of a narrow slit and obverse the subsequent correlated diffraction. In the degenerate SPDC process, one pump photon of energy $\hbar\omega$ turns into two spatially entangled photons of energy $\hbar\omega/2$, where $\omega$ is the angular frequency of the optical field. These correlations have been used to perform ghost imaging~\cite{pittman:95,howell:04,aspden:13} and, more recently, entangled field imaging~\cite{fickler:13}.

Under the assumption that walk-off inside the crystal is negligible, the two-photon SPDC mode function takes a simple form. As a function of the transverse wavevectors of photons \emph{A} and \emph{B}, $\bold{k_A}$ and $\bold{k_B}$ with $\bold{k}=k_x\bold{\hat{x}}+k_y\bold{\hat{y}}$, the two-photon mode function is given by 
\begin{align} \label{eq:SPDCsupp}
	\Phi_\text{Q}(\bold{k_A},\bold{k_B})=N\hspace{2pt}\tilde{E}\hspace{2pt}(\bold{k_A}+\bold{k_B})\hspace{4pt}\tilde{F}\left(\frac{\bold{k_A}-\bold{k_B}}{2}\right),
\end{align}
where $N$ is a normalization constant, $\tilde{E}(\bold{k})$ is the Gaussian-distributed angular spectrum of the pump laser,  and $\tilde{F}(\bold{k})$ is the phase-matching function \cite{pires:09}. In the paraxial wave approximation, the phase-matching function is of the form 
\begin{equation}\label{eq:Fsupp}
\tilde{F}(\bold{k})=\text{sinc}\left(\varphi+{L}\hspace{2pt}|\bold{k}|^2/{k_p}\right),
\end{equation} 
where $\varphi$ is the phase mismatch parameter, $L$ is the thickness of the crystal and $k_p$ is the wavevector of the pump inside the crystal. Since $|\varphi|\approx0$ in our experiment, we work in the regime where the phase--matching function can be approximated by a Gaussian distribution \cite{chan:07,padgett:12}. Further, we perform a Fourier transform to express the joint state in position space:
\begin{align}\label{eq:NFsupp} 
	\Psi_\text{Q}({\bold{r_B}},{\bold{r_A}})=N'\exp{\left(-\frac{\left|\bold{r_A}+\bold{r_B}\right |^2}{4 \sigma_p^2}\right)}\,\exp{\left(-\frac{\left|\bold{r_A}-\bold{r_B}\right|^2}{4\sigma_q^2}\right)},
\end{align}
where $\bold{r}=x\bold{\hat{x}}+y\bold{\hat{y}}$, $\sigma_p = 450 \mu$m is the $1/\text{e}^2$ width of the pump beam and $\sigma_q=\sqrt{L/2k_p} = 9.2\mu$m. The two-photon mode function is now separable in its vertical and horizontal components. We can thus integrate over the vertical degree of freedom and consider the joint mode from Eq.~(\ref{eq:NF}).

Since we have $\sigma_p \gg \sigma_q$, the term that governs the strength of the near-field correlations is $\exp[-(x_A+x_B)^2/(4\sigma_p^2)]$ while the term that mostly determines the mode of the singles is $\exp[-(x_A-x_B)^2/(4\sigma_p^2)]$, where $\sigma_p$ will be defined late (it is a function of $\sigma_q$). Indeed, if we assume perfect position correlations, $\delta ({x_A}-{x_B})$, the intensity profile of the singles is exactly given by the term $\exp[-(x_A+x_B)^2/(4\sigma_p^2)]$. \newline

\begin{figure}
	\centering
	\includegraphics[width=0.47\textwidth]{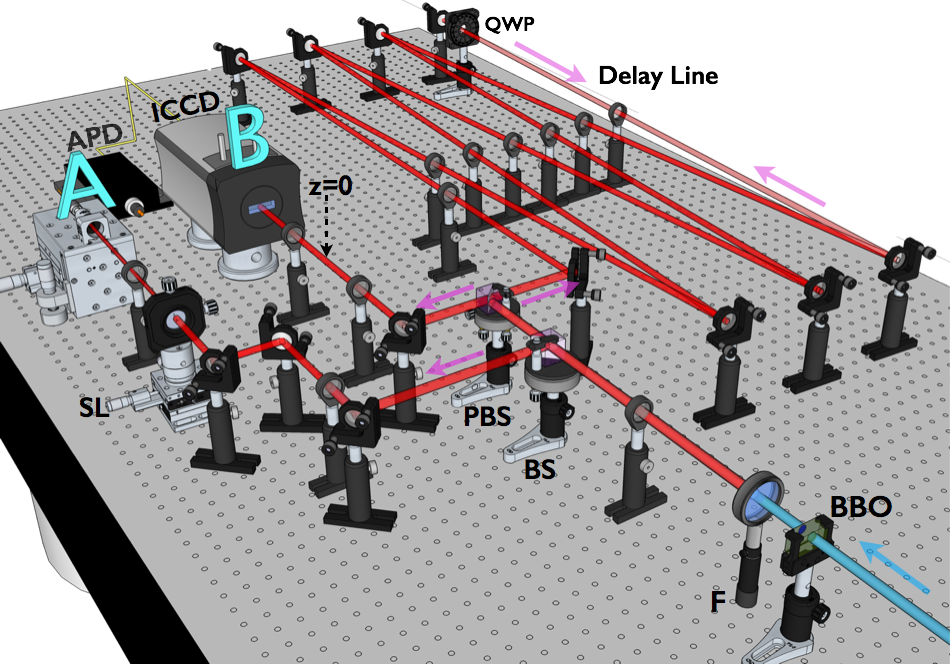} 
	\caption{\label{fig:quantumSUP} \textbf{Schematic of our implementation of Popper's experiment.} Entangled photon pairs are generated through SPDC at a $\beta$-barium borate (BBO) type-I  crystal. The pump beam is filtered out by a high-pass filer (F). The two photons of a given pair are separated by a 50/50 non-polarizing beam-splitter (BS). The mode of photon \emph{A} is imaged with unit magnification to a 10-$\mu$m-slit and then collected by a multimode optical fiber with a 200~$\mu$m core diameter that is connected to a single-photon detector that triggers the ICCD camera. Photon \emph{B}, initially vertically polarized, is reflected by the polarizing beam-splitter (PBS) towards an image-preserving delay line. Photon \emph{B} hits the final mirror of the delay line and makes its way back with a horizontal polarization because of its two passes in the quarter-wave plate. Photon \emph{B} thus traverses the PBS and reaches the ICCD camera, which records photon \emph{B} in the far-field of the BBO crystal. The BBO crystal is imaged with a magnification of one to the plane indicated by $z=0$~mm.} 
\end{figure}

\section{Experimental setup} 
In the experiment a frequency-tripled quasi CW mode-locked Nd-YAG laser (not shown) with a repetition rate of $100$~MHz and average output power of $P=150$~mW at $\lambda=355$~nm is used to pump a $3$~mm thick $\beta$-barium borate (BBO) crystal cut for type-I degenerate phase matching. The generated photon pairs (photon \emph{A} and photon \emph{B}) via SPDC are split out by means of a $50/50$ non-polarizing beam splitter (BS) and sent into an actual slit (path \emph{A}) and a delay line (path \emph{B}), respectively. Photon \emph{A} is imaged on a $10~\mu$m slit (SL) via a $4$f-system with a unit magnification, and then coupled into a $200~\mu$m core diameter multimode optical fibre. The coupled photons are detected by a silicon avalanche photodiode (APD) and used to trigger an intensified charge coupled device (ICCD) camera. On the other side, photon \emph{B} is sent to a delay line, which is made of a polarizing beam splitter (PBS), three $4$f-system and a quarter-wave plate (QWP). Photon \emph{B}, then, after a certain delay time, inquired by response time of APD and ICCD camera, is imaged on the ICCD camera, which was positioned \emph{i)} near-filed and \emph{ii)} far-field of the BBO crystal.   

The spatial modes \emph{A} and \emph{B} are imaged to the plane of the real slit and the ghost slit, respectively, with a unit magnification, such that the exponential term on the right of Eq.~(\ref{eq:NF}) determines the width of the correlations. Indeed, when photon \emph{A} is detected at position $x_A=0$, the  ghost slit takes the form of a Gaussian distribution.

We put a large aperture in the path of \emph{B} (near the far-field of the crystal), such that the phase-matching function can be approximated closely by a simple gaussian beam. The phase--matching function then becomes $\tilde{F}(\bold{k})=\exp(-k_p^2/(4\sigma_q^2))$ with $\sigma_q^2=\sigma_q^2+r_a^2$, where $r_a$ is related to the phase mismatch parameter $\varphi$ and, mostly, to the size of the aperture; $r_a$ needs to be measured. 

A  ghost image necessarily has finite resolution either limited by the numerical aperture of the lenses used or by the strength of the correlations. Here, the bottleneck is the latter.}



\begin{thebibliography}{99}

\bibitem{epr:35} 
A.~Einstein, B.~Podolsky, and N.~Rosen, {\em Can Quantum-Mechanical Description of Physical Reality Be Considered Complete?}, Phys. Rev. {\bf 47}, 777--780 (1935).

\bibitem{bell:64} 
J.~S.~Bell, {\em On the Einstein--Podolsky--Rosen paradox}, Physics {\bf 1}, 195--200 (1964).

\bibitem{ghirardi:88} 
G. C. Ghirardi, R. Grassi, A. Rimini, and T. Weber, {\em Experiments of the EPR type involving CP-violation do not allow faster-than-light communication between distant observers}, Europhys. Lett. {\bf 6}, 95--100 (1988).

\bibitem{popper:34} 
It is worth mentioning that we tested the latest version of the Popper's experiment, which was proposed in his 1982 book~\cite{popper:82} and known as Popper's experiment nowadays. The first version appeared in his 1934 publication (K.~Popper, {\em Zur Kritik der Ungenauigkeitsrelationen}, Naturwissenschaften {\bf 22}, 807--808 (1934)), but this version was later rejected by Popper as being incorrect.

\bibitem{popper:82} 
K. R. Popper, {\em Quantum theory and the schism in physics} (London: Hutchinson, 1982).

\bibitem{qureshi:05} 
T. Qureshi, {\em Understanding Popper's experiment}, Am. J. Phys. {\bf 73}, 541--544  (2005).

\bibitem{gerjuoy:06} 
E. Gerjuoy and A. M. Sessler, {\em Popper's experiment and communication}, Am. J. Phys. {\bf 74}, 643--648 (2006).

\bibitem{richardson:12} 
C. D. Richardson and J. P. Dowling, {\em Popper's thought experiment revisited}, Int. J. Quantum Inform. {\bf 10}, 1250033 (2012).

\bibitem{collett:87} 
M. Collett and R. Loudon, {\em Analysis of a proposed crucial test of quantum mechanics}, Nature {\bf 326}, 671--672 (1987).

\bibitem{kim:99} 
Y.-H. Kim and Y. Shih, {\em Experimental Realization of Popper's Experiment: Violation of the Uncertainty Principle?}, Found. Phys. {\bf 29}, 1849--1861 (1999).

\bibitem{short:01} 
A. J. Short, {\em Popper's experiment and conditional uncertainty relations}, Found. Phys. Lett. {\bf 14}, 275--284 (2001).

\bibitem{peres:02} 
A. Peres, {\em Karl Popper and the Copenhagen interpretation}, Studies In History and Philosophy of Science Part B:
Studies In History and Philosophy of Modern Physics {\bf 33}, 23--34 (2002).

\bibitem{ghirardi:07} 
G. Ghirardi, L. Marinatto, and F. de Stefano, {\em Critical analysis of Popper's experiment}, Phys. Rev. A {\bf 75}, 042107 (2007).

\bibitem{chan:07} 
K. Chan, J. Torres, and J. Eberly, {\em Transverse entanglement migration in Hilbert space}, Phys. Rev. A {\bf 75}, 050101 (2007).

\bibitem{pires:09} 
H. D. L. Pires and M. van Exter, {\em Near-field correlations in the two-photon field}, Phys. Rev. A {\bf 80}, 053820 (2009).

\bibitem{walborn:10} 
S. P. Walborn, C. Monken, S. P\'adua, and P. Souto Ribeiro, {\em Spatial correlations in parametric down-conversion}, Phys. Rep. {\bf 495}, 87--139 (2010).

\bibitem{bennink:02} 
R. S. Bennink, S. J. Bentley, and R. W. Boyd, {\em ``Two-photon'' coincidence imaging with a classical source}, Phys. Rev. Lett. {\bf 89}, 113601 (2002).

\bibitem{gatti:04} 
A. Gatti, E. Brambilla, M. Bache, and L. A. Lugiato, {\em Ghost imaging with thermal light: comparing entanglement and classical correlation}, Phys. Rev. Lett. {\bf 93}, 093602 (2004).

\bibitem{combourieu:92} 
M.-C. Combourieu, {\em Karl R. Popper, 1992: about the EPR controversy}, Found. Phys. {\bf 22}, 1303--1323 (1992).

\bibitem{pittman:95} 
T. Pittman, Y. Shih, D. Strekalov, and A. Sergienko, {\em Optical imaging by means of two-photon quantum entanglement}, Phys. Rev. A {\bf 52}, R3429 (1995).

\bibitem{howell:04} 
J. Howell, R. Bennink, S. Bentley, and R. Boyd, {\em Realization of the Einstein-Podolsky-Rosen Paradox Using Momentum- and Position-Entangled Photons from Spontaneous Parametric Down Conversion}, Phys. Rev. Lett. {\bf 92}, 210403 (2004).

\bibitem{aspden:13} 
R. S. Aspden, D. S. Tasca, R. W. Boyd, and M. J. Padgett, {\em EPR-based ghost imaging using a single-photon-sensitive camera}, New J. Phys. {\bf 15}, 073032 (2013).

\bibitem{fickler:13} 
R.~Fickler, M.~Krenn, R.~Lapkiewicz, S.~Ramelow, and A.~Zeilinger, {\em Real-time imaging of quantum entanglement}, Sci. Rep. {\bf 3}, 1914 (2013).

\bibitem{padgett:12} 
M.~P.~Edgar, D.~S.~Tasca, F.~Izdebski, R.~E.~Warburton, J.~Leach, M.~Agnew, G.~S.~Buller, R.~W.~Boyd, and M.~J.~Padgett, {\em Imaging high-dimensional spatial entanglement with a camera}, Nat. Commu. {\bf 3}, 984 (2012).

\end{thebibliography}
\end{document}